\documentclass[11pt]{article}

\makeatletter 

\usepackage{color}
\usepackage{amsmath,amssymb}
\usepackage{wrapfig,url,slashed}
\usepackage[hypertex]{hyperref}    
\usepackage{graphicx,cite}
\renewcommand\citepunct{,\penalty\@M\hskip.13emplus.1emminus.1em\relax} 
\usepackage{enumerate}

\newcommand{\dd}{{\mathrm d}}      

\newcommand\Order{\mathop{\mathcal{O}}}

\newcommand{\sgn}{\mathop{\mathrm{sgn}}}

\newcommand{\Br}{\mathop{\mathrm{Br}}}

\newcommand{\vc}[1]{{\boldsymbol #1}} 

\newcommand{\@diff}   [4]{\dfrac{#4#3#1}{#4#2#3}}
\newcommand{\diff}    [2]{\@diff{#1}{#2}{}{\dd}}
\newcommand{\pdiff}   [2]{\@diff{#1}{#2}{}\partial}
\newcommand{\fdiff}   [2]{\@diff{#1}{#2}{}{\delta}}
\newcommand{\ndiffnum}[3]{\@diff{#1}{#2}{^#3}\dd}
\newcommand{\npdiff}  [3]{\@diff{#1}{#2}{^#3}\partial}
\newcommand{\@difftwo}[4]{\dfrac{#4^2#1}{#4#2\,#4#3}}
\newcommand{\difftwo} [3]{\@difftwo{#1}{#2}{#3}{\dd}}
\newcommand{\pdifftwo}[3]{\@difftwo{#1}{#2}{#3}{\partial}}

\newcommand{\un}[1]{{\mathrm{\,#1}}} 
\newcommand{\TeV}{\un{TeV}}
\newcommand{\GeV}{\un{GeV}}
\newcommand{\MeV}{\un{MeV}}
\newcommand{\keV}{\un{keV}}

\def\@xxxEV{\@ifnextchar-{\@xxxEV@minus}{\@xxxEV@plus}}
\def\@xxxEV@plus#1#2{%
  \ifnum{#1=0}{}\else\ifnum{#1=1}{10}\else {10^#1}\fi\fi #2}
\def\@xxxEV@minus#1#2 {10^{-#1}{\rm\,#2}}

\newcommand{\TEV}[1]{\@xxxEV{#1}{\TeV}}
\newcommand{\GEV}[1]{\@xxxEV{#1}{\GeV}}
\newcommand{\MEV}[1]{\@xxxEV{#1}{\MeV}}
\newcommand{\KEV}[1]{\@xxxEV{#1}{\keV}}

\newcommand{\lrfp}[3]{ \left(\frac{#1}{#2} \right)^{#3}}

\def\EE{\@ifnextchar-{\@@EE}{\@EE}}
\def\@EE#1{\ifnum#1=1\times10\else\times10^{#1}\fi}
\def\@@EE#1#2{\!\times\!10^{-#2}}

\def\T{\@ifnextchar^{\T@u}{\@ifnextchar_{\T@d}{}}}
\def\T@u^#1{{^{#1}}\T}
\def\T@d_#1{{_{#1}}\T}

\let\lsim\lesssim
\let\gsim\gtrsim
\newcommand{\beq}{\begin{equation}}
\newcommand{\eeq}{\end{equation}}
\newcommand{\bea}{\begin{eqnarray}}
\newcommand{\eea}{\end{eqnarray}}
\newcommand{\pmat}[1]{\begin{pmatrix}#1\end{pmatrix}}

\newcommand{\s}[1]{_\mathrm{#1}}    

\newcommand{\suprm}[1]{^\mathrm{#1}} 

\newcommand{\TO}{\,\text{--}\,}

\let\tilde\widetilde 

\newcommand{\Hu}{H\s u}

\newcommand{\bU}{{\bar U}} \newcommand{\bD}{{\bar D}} \newcommand{\bE}{{\bar E}}

\newcommand{\stau}{\tilde{\tau}}

\newcommand{\PT}{P\s T}
\newcommand{\ET}{E\s T}
\newcommand{\invfb}{\un{fb^{-1}}}

\makeatother 

\begin{document}
\baselineskip=18pt

\begin{titlepage}

\begin{flushright}
UT--11--06\\
IPMU--11--0033\\
\end{flushright}

\vskip 1.35cm
\begin{center}
{\Large \bf
Stau Kinks at the LHC
}
\vskip 1.2cm
Shoji Asai$^{1}$, Yuya Azuma$^{1}$,
Motoi Endo$^{1,2}$, Koichi Hamaguchi$^{1,2}$, Sho Iwamoto$^{1}$
\vskip 0.4cm

{\it $^1$  Department of Physics, University of Tokyo,
Tokyo 113--0033, Japan\\
$^2$ Institute for the Physics and Mathematics of the Universe,
University of Tokyo,\\ Chiba 277--8568, Japan
}

\vskip 2cm

\abstract{
The kink signature of charged tracks is predicted in some SUSY models,
and it is very characteristic signal at collider experiments.
We study the kink signature at LHC using two models, SUSY models with a gravitino LSP 
and a stau NLSP, and R-parity violating SUSY models with a stau (N)LSP.
We find that a large number of kink events can be discovered in a wide range of the SUSY parameters,
when the decay length is $\Order (10 \TO 10^5)\un{mm}$.
Model discrimination by identifying the daughter particles of the kink tracks is also discussed.
}
\end{center}
\end{titlepage}

\setcounter{footnote}{0}
\setcounter{page}{2}

\section{Introduction}

In most of supersymmetric (SUSY) standard models, it is assumed that the R-parity is exactly conserved, and
the lightest SUSY particle (LSP) is the lightest neutralino. The conservation of the R-parity makes the LSP neutralino
stable and promotes it to a viable candidate for the dark matter.
A characteristic signature for such a scenario at the Large Hadron Collider (LHC)
is a large missing transverse momentum.

However, some SUSY models provide not such a signature but more exotic ones.
A popular example is the SUSY models with a gravitino LSP, such as in gauge-mediated SUSY breaking models~\cite{Giudice1999twg}.
In this case, the next-to-lightest SUSY particle (NLSP) becomes long-lived due to the extremely weak interaction of the gravitino.
Another example is the R-parity violation\cite{Barbier2005rvs}.
Recently some of the authors studied cosmological constraints on the R-parity violation
taking into account the effect of lepton flavor violations, and showed that the baryon asymmetry of the universe favors very tiny R-parity violations~\cite{Endo2009lfv}. This means that the LSP (or NLSP, if the LSP is the gravitino) decays into Standard Model particles with a long decay length.

The long-lived superparticles in such scenarios provide various kinds of interesting signatures at the LHC~\cite{Fairbairn2007Smp,Feng:2010ij}.
The signatures crucially depend on the decay length $c\tau$ and the electric charge of 
the long-lived superparticle $X$:
\begin{itemize}
\item[(i)] Stable signature:
For $c\tau\gsim L$, where $L\simeq \Order({\rm m})$ is a typical detector size,
most of the produced $X$ particles pass through the detector before decaying. 
If $X$ is electrically neutral, the signature is a missing transverse momentum, 
which is similar to the conventional SUSY scenario. 
However, when $X$ is electrically charged, as in the case of long-lived stau scenario, 
it provides an anomalous track of large ionization and/or low velocity~\cite{CHAMPsignals}.
Furthermore, the charged track can be used to determine the masses of $X$~\cite{Nisati:1997gb,Hinchliffe:1998ys,Ambrosanio2000MSB,Ellis:2006vu} and other superparticles~\cite{Hinchliffe:1998ys,Ellis:2006vu,other_masses},
as well as the spin and other properties of superparticles~\cite{spin_etc}.
The lifetime of $X$ may also be measured by looking for its delayed decay after being stopped~\cite{lifetime:delayed_decay}.
Recently CMS has reported the lower limit of the stable/stopped gluino mass,
398/370~GeV respectively, using the highly ionized track and 
empty bunch~\cite{cms1,cms2}.
ATLAS also reported lower limit on the stable gluino mass, 562~GeV~\cite{Aad2011ssh}.
  
\item[(ii)]  In-flight decay:
When the decay length satisfies $\Order({\rm cm})\lsim c\tau \lsim N L$, where $N$ is the total number of $X$ produced through the experiment, its in-flight decays can be seen within the detectors~\cite{Ambrosanio2000MSB,Kawagoe:2003jv,Asai2008ToA,longlived_LHC,Asai2009MMo}.\footnote{The probability of the $X$ particle decaying within the distance $L$ is given by $P(L)=1-\exp (-L/\beta\gamma c \tau)$, where $\beta$ is the velocity of $X$ and $\gamma=(1-\beta^2)^{-1/2}$, and typically $\beta\gamma = \Order(1)$. For $c\tau \gg L$, it is evaluated as $P(L) \simeq L/\beta\gamma c\tau \simeq L/c\tau$, and hence the number of in-flight decay events is given by $\sim (L/c \tau)N$.}
In the case of neutralino NLSP decaying into gravitino LSP, non-pointing photons may be observed~\cite{Kawagoe:2003jv}.
If $X$ is charged, its in-flight decay can cause a disappearing track~\cite{Asai2008ToA,Asai2009MMo} or a kinked charged track.
The lifetime measurement may also be possible by studying the distribution of, or simply by counting the number of, these in-flight decay events~\cite{Ambrosanio2000MSB,Kawagoe:2003jv,Asai2008ToA,longlived_LHC}.
\end{itemize}
Note that these two cases are not always mutually exclusive; for $L\lsim c\tau \lsim N L$, both types of signatures can be studied simultaneously.

In this paper, we will study the latter case (ii),  in particular, the kinks of charged tracks which are produced when long-lived charged particles decay into another charged particle inside the detector. 
Specifically, we investigate the feasibility of probing the SUSY models by analyzing such kink tracks 
in the transition radiation tracker (TRT) of the ATLAS detector,
which provides us continuous detection of charged tracks.
As underlying models, we concentrate on the following two models; (a) SUSY models with a gravitino LSP and a stau NLSP, and (b) R-parity violating SUSY models with a stau (N)LSP.
We show that, in a realistic setup, a large number of kink events can be discovered in a wide range of the SUSY parameters.
We also discuss the possibility of model discrimination by identifying the daughter particles of the kink tracks.

\section{Kink Tracks}
\label{sec:kink-atlas-detector}

A kink track signature and possible backgrounds are discussed here. 
When a particle decays in tracking detectors and produces one charged daughter particle,
the track would bend abruptly, which forms a kink track.
In the ATLAS detector, charged tracks are observed in the inner detector,
which consists of the pixel detector, the semi-conductor tracker (SCT) and the
TRT, located from the inside to outside of the beam axis respectively.
In particular, the TRT, which is a continuous tracking detector, is suited to observe the kinks
directly.

The TRT detector \cite{AtlasTDRInner1} consists of the barrel part and the end-cap part.
The barrel part covers the collision point cylindrically: $|z|<712\un{mm}$ and $563\un{mm}<R<1066
\un{mm}$, while the end-cap part caps both sides.\footnote{%
Our coordinate system follows the ATLAS standard:
$z$ is along the beam axis whose origin is the nominal collision point,
$R$ is the distance from the beam axis, and $\theta$ and $\phi$ is the azimuthal and the polar angles.}
The barrel part consists of three modules: the first module (the innermost one) sits at the range of
$563\un{mm}<R<694\un{mm}$, the second one at $697\un{mm}<R<860\un{mm}$, and the third
one (the outermost one) at $863\un{mm}<R<1066\un{mm}$.
Each module contains straw tubes, which are aligned parallel to the beam axis and provide
$R$\,-\,$\phi$ information of the track position.
The barrel TRT is composed of 73 layers of the straws in total. 
When a charged particle passes through the TRT, 
it hits a large number (typically $\sim 40$) of the straws.
In the analysis we use the barrel part of the TRT
to measure the kink tracks.

\begin{figure}[t]
 \begin{center}
  \includegraphics[scale=0.8]{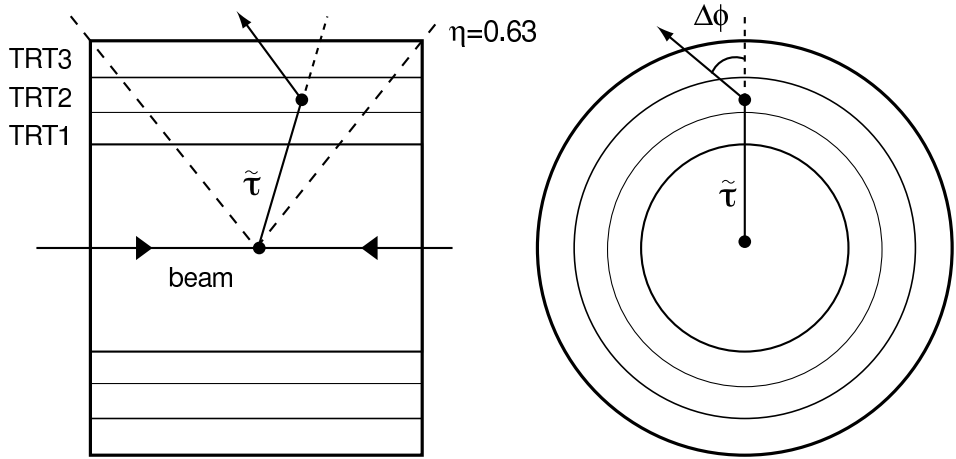}
  \caption{Schematic pictures of the decay which generates a kink track in the TRT.
  In the left panel, the tracks are projected onto the plane which includes the beam axis.
  In the right panel, they are projected onto the plane which is perpendicular to the beam axis.}
  \label{fig:detector}
 \end{center}
\end{figure}

The kink track is identified by reconstructing both the tracks of the charged mother and daughter particles.
Schematic pictures of the kink track are shown in Fig.~\ref{fig:detector}.
The mother particle flies through the pixel and the SCT, and reaches the TRT.
Consider that the mother particle decays in the TRT. The track terminates halfway,
but is reconstructed very well by the information from the pixel detector and
the SCT~\cite{Asai2009MMo}. A good momentum
resolution~\cite{ATLAS2008TAE,Asai2009MMo} is also expected, 
and it is improved by combining the information from the
TRT~\cite{ATLAS2008epa}. Moreover, it may be possible to measure the lifetime and the mass of
the mother particle (or their combination) by the information on the momentum, the time-of-flight and
the decay length, depending on the models~\cite{Asai2008ToA,Asai2009MMo}.

The track of the mother particle terminates when the particle decays.
As we will require the mother particle to be $|\eta| < 0.63$, where $\eta$ is the pseudorapidity, it would pass through the full volume of the 3rd module of the TRT if it does not decay.
So we can recognize the termination of the track by identifying the absence of the TRT hits (less than noise level) after the decay point.
Moreover, a number of hits will be observed in the 3rd module for the daughter particle(s).
However, when the daughter particle has the same azimuthal angle as the mother, these two tracks can not be distinguished in the TRT.
In order to reconstruct the tracks separately to observe a kink, the kink is required to be $\Delta\phi > 0.1$, 
where the kink angle $\Delta\phi$ is defined as the difference of the azimuthal angles between
the mother and daughter tracks. 
The charged daughter particle is also required not to escape from the end-cap of the TRT, which guarantees that
the daughter particle passes through full volume of the TRT 3rd module (about 15 hits are expected),
and  the daughter track can be reconstructed with reasonable efficiency.
Using the simulations based on G4, in which all detector materials are taken into account, 
the reconstruction efficiency is estimated to be 70\% for $\PT > 20\GeV$, and 60\% even for $\PT\sim10\GeV$~\cite{gakkai}.
This reconstruction efficiency is taken into account in this study. 

We also need to comment that it is difficult or impossible to reconstruct
the tracks of soft~\cite{ATLAS2008epa,Asai2009MMo} or neutral daughter particles.
Especially we can hardly reconstruct the kink for the case where the charged daughter particle is soft, which usually occurs when a neutral, i.e.~missing, daughter particle is almost degenerate with the mother particle.
Even in such cases, the mother track can be identified  as a disappearing track and we can measure the lifetime of the mother particle~\cite{Asai2009MMo} .
 In this paper we consider models in which the decay products are much lighter than the mother particle.
 We can identify not only the mother but also the daughter particles.

It is possible to identify the particle species of the daughter track from the information of various detectors. 
When the daughter particle is an electron, we obtain signals in the TRT and
the electromagnetic (EM) calorimeter. 
When a muon, a characteristic signature arises in the muon spectrometer. 
When the mother particle decays into jets, they are identified by the EM and
hadronic calorimeters. 
In these three cases, the momentum of the daughter can be determined.
Lastly, when the daughter is a tau lepton, it decays right after the production
and emits an electron, a muon, or hadrons.
The ratio of the kink events with the $e$/$\mu$/jets
final state would enable us to identify the daughter tau production statistically, which will be discussed in Sec.~\ref{sec:model_discrimination}.
Once the decay branching fraction of the mother particle is known, it helps us to discriminate
the underlying models, such as the models with the R-parity violation or
with a gravitino LSP.

The masses of the mother particles and the invisible daughter can also be determined.
Velocity of the mother particle can be measured with an accuracy of $\Order (10)\%$
from pixel energy loss information~\cite{Aad2011ssh}, and accuracy of $\Order (1)\%$ from the timing information
in the hadronic calorimeter and the muon detectors~\cite{Ambrosanio2000MSB,ATLAS2008epa}. 
If the typical $c\tau \gamma$ is longer than $\Order (1)\un{m}$,  the mother particle's mass 
can be measured precisely. 
The transverse momentum $\PT$ of the visible daughter particles can be measured in TRT 
and the kinematic edge of the $\PT$ distribution carries the information of the invisible daughter particle's mass, with a 
typical resolution of $\Order (1)\GeV$.

The possible background sources for  the kink signature are as follows:

\begin{enumerate}

\item In-flight decays of hadrons, such as $\pi^{\pm}$ and $K^{\pm} $, have similar topology. 
However, it is considered that the hadronic background events are suppressed by requiring a high transverse 
momentum of the mother particle and a large kink angle. 
This is because a boosted particle is likely to emit the decay products forward. 
Consider a charged particle $X$ decaying into a charged particle $A$ and another (neutral) particle.
The kink angle $\Delta\phi$ (the difference of the azimuthal angles between $X$ and $A$)
has a kinematical bound
\begin{equation}
 |\sin\Delta\phi|
 < \frac{1}{\sin \theta_X} \frac{m_X |\vc p_A\suprm{CM}|}{m_A |\vc p_X\suprm{Lab}|}
 < \frac{1}{\sin \theta_X} \frac{m_X^2/2}{m_A \PT(X)},
\end{equation}
where ``Lab'' denotes the laboratory frame, and ``CM'' does the rest frame of $X$. Here, $m_X$ and
$m_A$ are the masses of $X$ and $A$, respectively. The 3-momentum of the particle $i$ is represented
by $\vc p_i^F$ in the frame $F$, while $\PT(X)$ shows the transverse momentum of $X$ in
the laboratory frame. The angle $\theta_X$ is the polar angle of $X$, that is, the angle between $X$
 and the beam axis. In the analysis, we consider the decay of the mother particle to be
restricted in $|\eta| < 0.63$, which corresponds to $0.83 < \sin \theta_X < 1$. In order to suppress the
in-flight decay background, we impose $\PT(X) > 100\GeV$ and $\Delta\phi > 0.1$.
Using the simulation based on {\tt PGS}, the contribution of in-flight decay is estimated to be negligibly small.  

\item Stable charged hadrons ($\pi^\pm$, $K^\pm$, $p$, etc.) interact hadronically with materials of the inner detectors and
the direction of the particle is changed significantly.
This process also makes kink signatures.
However, background events with stable charged particles are also suppressed by the above requirement that the mother particle must have $\PT>100\GeV$.
\item The unphysical tracks produced by noise hits in the pixel and SCT detectors can have $\PT > 100\GeV$,
but these contributions can be suppressed dramatically by requiring the daughter track.

\end{enumerate}
Background rate is expected to be very low after all selections, which we will discuss later, are applied.

\section{Models}

In this section we briefly introduce supersymmetric models which can produce kink tracks at the LHC.
We also discuss the model discrimination in Sec.~\ref{sec:model_discrimination}.

\subsection{Models with a gravitino LSP and a stau NLSP}
\label{sec:models-with-grav}

In supersymmetric models with a gravitino LSP and a stau NLSP, under the R-parity conservation,
the stau decays into a tau lepton and a gravitino and becomes long-lived.
Its lifetime is given by
\begin{eqnarray}
c\tau &\simeq&
\left(\frac{m_{\stau}^5}{48\pi M_p^2 m_{3/2}^2}\right)^{-1}
\nonumber\\
&\simeq& 550\un{mm}
\left(\frac{m_{\stau}}{200\GeV}\right)^{-5}
\left(\frac{m_{3/2}}{1\keV}\right)^2\,,
\label{eq:ctaugravitino}
\end{eqnarray}
where $M_p$ is the reduced Planck scale, $m_{\stau}$ is the stau mass,
 and $m_{3/2}$ is the gravitino mass.
As we shall see, the stau kink may be observed at the TRT 
for $m_{3/2}\sim \Order(0.01\TO1)\keV$.
Interestingly, this corresponds to the most challenging gravitino mass range from 
cosmological point of view~\cite{gravitino_keV,Feng:2010ij}, which will be discussed in Sec.~\ref{sec:discuss}.
The produced tau lepton decays promptly, either leptonically or hadronically.
Then, the signature will be a kink track with the daughter particle being an electron, a muon, or a tau-jet.

The stau can be long-lived also in the case of an axino LSP and a stau NLSP. In the hadronic, or KSVZ, 
axion models~\cite{KSVZ}, the stau lifetime becomes much longer than the detector size~\cite{Brandenburg:2005he}, and the heavy charged particles will be observed in the muon system. 
In the alternative DFSZ axion models~\cite{DFSZ}, the stau decay length can be  as short as $\Order(1)\un{m}$~\cite{Martin:2000eq}.
 In such a case, the stau decay into an axino and a tau lepton may leave a kink signature. Hereafter, 
 we do not discuss the axino LSP case for simplicity, keeping in mind that the models with the axino LSP in the DFSZ axion models can be studied in the same way as the gravitino LSP case.

\subsection{R-parity Violation and Cosmological Constraints}
\label{sec:r-parity-violation}

If the R-parity is violated, the following terms are allowed in the superpotential:
\begin{equation}
 W\s{RpV}
  = \frac12\lambda_{ijk} L_i L_j \bE_k
  +  \lambda'_{ijk} L_i Q_j \bD_k
  + \kappa_i L_i \Hu
  + \frac12\lambda''_{ijk}\bU_i \bD_j \bD_k,\label{eq:RpV-Superpotential}
\end{equation}
where we use the conventions, $\lambda_{ijk}=-\lambda_{jik}$ and $\lambda''_{ijk}=-\lambda''_{ikj}$.
The first three terms violate the lepton number $L$, while the last term does the baryon number $B$.

Unless the lepton flavor violations are extremely suppressed, the $B$- or $L$-violation would erase the existing baryon asymmetry before the electroweak transition.
Therefore, once we assume that the current baryon asymmetry is generated before the electroweak transition, the R-parity violations are severely constrained.
The constraints are described as~\cite{Endo2009lfv},
\begin{align}
 \sqrt{\sum_{ijk}|\lambda_{ijk}|^2}   &\lesssim (0.6 \TO 1) \EE-6\,, \label{eq:boundLLE}\\
 \sqrt{\sum_{ijk}|\lambda'_{ijk}|^2}  &\lesssim (3   \TO 6)   \EE-7\,, \label{eq:boundLQD}\\
 \sqrt{\sum_{ijk}|\lambda''_{ijk}|^2} &\lesssim (4   \TO 5)   \EE-7\,, \label{eq:boundUDD}\\
 \sqrt{\sum_{i}\left|\frac{\kappa_i}{\mu} \right|^2} &\lesssim (1 \TO 2)\EE-6
 \lrfp{\tan\beta}{10}{-1}\,,
 \label{eq:boundLHu}
\end{align}
for the squark mass $m_{\tilde q} \simeq 200 \TO 1200\GeV$ and the slepton mass $m_{\tilde l}
\simeq 100 \TO 400\GeV$. Here, $\mu$ denotes the Higgsino mass parameter, and $\tan\beta$
is the ratio of the up- and down-type Higgs vacuum expectation values.
The above severe constraints imply that,
if R-parity is violated, the lightest superparticle among the superpartners of the Standard Model particles (which we call MSSM-LSP\footnote{%
If gravitino is the LSP, the NLSP is the MSSM-LSP. If not, the LSP is the MSSM-LSP.
}) becomes long-lived and may leave interesting signatures at the LHC.
We assume that the lighter stau is the MSSM-LSP and decays to Standard Model particles via the R-parity violating interactions.

In our analysis the R-parity is considered to be violated only in the term $\frac12\lambda_{ijk}L_i L_j
\bE_k$; we assume the other sectors including the SUSY breaking terms have no R-parity violation.
Then, the stau decays into 2- or 4-leptons. The decay channels for each type of the R-parity violating
coupling are summarized as follows.
\begin{description}
 \item [$\underline{ \lambda_{123}}$]:
             The decay rate of stau is
             $\Gamma\s{tot}=\dfrac{(\lambda\sin\theta)^2}{8\pi}m_{\stau}$, with
             $\Br(\stau\to e\nu)=\Br(\stau\to\mu\nu)=0.5$.
 \item [$\underline{\lambda_{i33}}$ $(i=1,2)$]:
             The decay rate is $\Gamma\s{tot} = \dfrac{\lambda^2(1+\sin^2\theta)}{16\pi}m_{\stau}$,
             with $\Br(\stau\!\to\!\tau\nu) = \dfrac{1}{1+\sin^2\theta}$ and~
             $\Br(\stau\!\to\!l_i\nu) = \dfrac{\sin^2\theta}{1+\sin^2\theta}$.
 \item [$\underline{ \lambda_{i3k}}$ $(i=1,2;\ k=1,2)$]:
             Here is only one decay mode:
             $\Gamma\s{tot}=\Gamma(\stau\to l_k\nu)=\dfrac{(\lambda\cos\theta)^2}{16\pi}m_{\stau}$.
 \item [$\underline{ \lambda_{121}, \lambda_{122}}$]:
             Stau decays into 4-body.
\end{description}
Here, we ignore the mass of the leptons.
$\lambda$ is the corresponding coupling constant, and $\nu$ denotes
both neutrinos and anti-neutrinos.
$\theta$ is the mixing angle of stau, which is defined as
\begin{equation}
 \pmat{\stau_1\\\stau_2}
= \pmat{\cos\theta&\sin\theta\\-\sin\theta&\cos\theta}\pmat{\stau\s L\\\stau\s R},
\end{equation}
where $\stau_1$ and $\stau_2$ are mass eigenstates, and $\stau_1$ is now assumed to be the MSSM-LSP.
Hereafter, for simplicity, we do not discuss the case where stau decays into 4-body,
although it may be analyzed in a similar way as the 2-body case.

For later convenience, we define an effective coupling $\lambda\s{eff}$, so that the decay rate becomes
\begin{equation}
 \Gamma\s{tot} = \frac{\lambda\s{eff}^2}{16\pi}m_{\stau}.
\end{equation}
That is,
\begin{equation}
 \lambda\s{eff} =
\begin{cases}
 \lambda_{123}\sqrt2\sin\theta
      &\text{for $\lambda_{123}$,}\\
 \lambda_{i3k}\cos\theta
      &\text{for $\lambda_{131}, \lambda_{132}, \lambda_{231}, \lambda_{232}$,}\\
 \lambda_{i33}\sqrt{1+\sin^2\theta}
      &\text{for $\lambda_{133}, \lambda_{233}$.}
\end{cases}
\end{equation}
Then the decay length $c\tau$ of stau is estimated as
\begin{equation}
 c\tau = c\left(\frac{\lambda^2\s{eff}}{16\pi}m_{\stau}\right)^{-1} \simeq 990\un{mm}\lrfp{\lambda\s{eff}}{1\EE-8}{-2}\lrfp{m_{\stau}}{100\GeV}{-1}.
\label{eq:ctauRpV}
\end{equation}
Therefore, if the coupling is $\lambda\s{eff} \lsim \Order (10^{-8})$, 
which satisfies the cosmological upper bound 
Eq.~(\ref{eq:boundLLE}),
the stau decay length is longer than $\Order(10)\un{mm}$.

Such small R-parity violation is cosmologically favored in the gravitino LSP scenario~\cite{Takayama2000Gdm,Buchmuller2007gdm}. When the 
gravitino mass is $m_{3/2} > 5\GeV$, the R-parity violation of $10^{-14} < \lambda_{ijk}, 
\lambda'_{ijk} < 10^{-7}$ naturally reconciles three well-motivated paradigms~\cite{Buchmuller2007gdm}: 
the primordial nucleosynthesis, gravitino dark matter, 
and the thermal leptogenesis.
On the other hand, the relic gravitino decays via the R-parity violation, which may be observed 
by cosmic rays~\cite{Takayama2000Gdm,Buchmuller2007gdm,GraviCosmi}.
For $m_{3/2} > 100\GeV$, 
the R-parity violation is constrained so tightly by cosmic ray measurements 
that the stau decay length becomes typically much longer than the detector size.
For $m_{3/2}\sim 10\GeV$, $\lambda_{ijk} = \Order(10^{-8})$ can satisfy the 
cosmic-ray bounds, and the stau kinks can be observed.

In our analysis we mainly concentrate on the $L_i L_j \bE_k$ interaction for simplicity.
When the R-parity is violated in the bilinear $L_i\Hu$ term,
the stau decay causes a kink track, or splits into jet tracks, in the TRT,
depending on the pattern of R-parity violation in the superpotential as well as
in the soft terms.
The former case can be described as the same picture
as the $\lambda_{ijk}$ case and thus covered by this paper.
The signature of the latter case is a few jets plus 0--1 lepton, which is out
of our scope and needs another method.

\subsection{Model Discrimination}
\label{sec:model_discrimination}

Once the mother particle of the kink track
is known, e.g.~by studying decay chains of the SUSY events, the underlying model is shed light on by
studying the daughter particle as well as measuring the lifetime and the mass spectrum. In particular,
each model predicts a characteristic pattern of the branching ratio of the decay channels. Actually, when
the R-parity is violated, the violation parameter determines decay products such as $e$, $\mu$, $\tau$ and
jets. When the stau decays into the gravitino, the tau leptons are produced. Since $\tau$
decays into other leptons or hadrons, we expect to observe $e$, $\mu$ and jets at specific number ratios.
Therefore, the reconstruction of the daughter particle is crucial for revealing the kink model.
To be explicit, we summarize the branching ratios of the lepton production channels and the relative
amounts of the measurable final-state particles for several models in Tab.~\ref{tab:DaughterFractions}.

\begin{table}
\begin{center}
\caption{%
The expected fractions of the daughter particles.
The number ratios of $e, \mu$ and $\tau$ produced at the $\stau$ decay are given in the second
group of the columns for each models. The particles/jet in the third group are observed in detectors,
where the percentage shows the fraction of the number of each event.
We use $\sin\theta = 1$ numerically for $\lambda_{133}$ and $\lambda_{233}$.}
\label{tab:DaughterFractions}\vspace{6pt}
\begin{tabular}{|c||c@{\ :\ }c@{\ :\ }c||r|r|r|}
\hline
 \multicolumn{1}{|c||}{Models}
 & $e$ & $\mu$ & \ \ $\tau$\ \ \
 & \multicolumn{1}{|c|}{$e$} & \multicolumn{1}{|c|}{$\mu$} & \multicolumn{1}{|c|}{$\tau$-jet}\\\hline
Gravitino LSP   & 0 & 0 & 1        &  18\%   &  17\%   &   65\% \\\hline
$\lambda_{123}$ & 1 & 1 & 0        &  50\%   &  50\%   &  \multicolumn{1}{|c|}{$-$}  \\\hline
$\lambda_{i31}$ & 1 & 0 & 0        & 100\%   & \multicolumn{1}{|c|}{$-$} & \multicolumn{1}{|c|}{$-$} \\\hline
$\lambda_{i32}$ & 0 & 1 & 0        & \multicolumn{1}{|c|}{$-$}  & 100\%  & \multicolumn{1}{|c|}{$-$}  \\\hline
$\lambda_{133}$ & $\sin^2\theta$&0&1 &  59\%   &   9\%   &   32\% \\\hline
$\lambda_{233}$ & 0&$\sin^2\theta$&1 &   9\%   &  59\%   &   32\% \\\hline
\end{tabular}
\end{center}
\end{table}

\begin{table}[t]
\begin{center}
\caption{%
mSUGRA parameters at the BC1 benchmark point.
We will vary the R-parity violating couplings; since we consider very tiny R-parity violation, they do not affect the mass spectrum.
}
\label{tab:SUSYparams}\par
 \begin{tabular}[t]{|c|c|c|c|c|}
\hline
$M_0$   & $M_{1/2}$  & $\tan\beta$ & $A_0$ & $\sgn\mu$
\\\hline
$0\GeV$ & $400\GeV$  & $13$        & $0\GeV$ & +
\\\hline
\end{tabular}
\end{center}
\end{table}

\begin{figure}[t]
 \begin{center}
  \includegraphics[width=240pt]{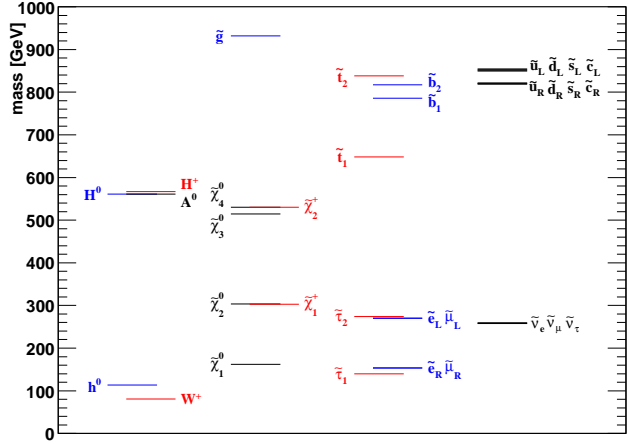}
  \caption{The mass spectrum of our model point at the weak scale.}
  \label{fig:BC1spectrum}
 \end{center}
\end{figure}

\section{Analysis}\label{sec:analysis}
We analyze the kink signature in the LHC by the Monte Carlo simulation.
We consider the following three LHC setups: the center-of-mass energies are $E\s{CM}=7\TeV$,
$8\TeV$, and~$14\TeV$, with the integrated luminosity $2\un{fb}^{-1}$, $5\un{fb}^{-1}$ and
$10\un{fb}^{-1}$, respectively.
The high $\PT$ one jet plus large missing $\ET$ trigger is used, which is standard for the SUSY searches.
As described later,  $\PT >120\GeV$  is required for the leading jet, 
and the missing $\ET $ is required to be larger than $100\GeV$
at the off-line analysis.  
These selections are related to the trigger and the trigger efficiency is higher than 90\% for
the events satisfying these conditions. 
The similar trigger condition will be used for the luminosity of $10^{32}$ to $10^{33}$ $\un{cm^{-2}s^{-1}}$
for the high $\PT$ one jet plus large missing $\ET$ trigger.
The single jet trigger is also useful for this analysis, and
we can make use of the trigger redundancy.


The benchmark point BC1~\cite{Allanach2007msi} is used with R-parity
violating couplings varied.\footnote{%
Although the gravitino LSP scenario, one of our concern, is usually realized in the GMSB framework,
we use this benchmark point in the analysis.
The result is expected to be 
insensitive to the detail of the mass spectrum.}
The parameters are summarized  in Tab.~\ref{tab:SUSYparams}
and the mass spectrum at the weak scale is shown in  Fig.~\ref{fig:BC1spectrum}.
The MSSM-LSP is the lighter stau with a mass of $140\GeV$.
Note that the tiny R-parity violation, one of our concern, hardly affects to the mass spectrum.
The SUSY events are generated mainly via the squark-squark production channel and
the squark-gluino channel; especially, the former dominates at the $7\TeV$ LHC, while in
the other setups they are in the same order.
We utilize {\tt SUSY-HIT 1.3}~\cite{SUSYHIT} to calculate
the SUSY mass spectrum and the decay table.
The SUSY events are generated by {\tt PYTHIA 6.4.23}~\cite{Pythia6.4},
and the detector simulation relies on {\tt PGS4}~\cite{PGS4web}.
In addition, we use {\tt TAUOLA 2.9}\cite{TAUOLA2.4}
(combined with {\tt TAUOLA C++ Interface}~\cite{TAUOLACPP}) to simulate the decay of tau leptons,
and {\tt PYTHIA} to successive decays of the $\pi^0$ and $\eta$ mesons.

For the cases where the stau emits $e$ or $\mu$, the following selections are applied.
\begin{enumerate}
 \item At least one jet with $\PT>120\GeV$ and the missing $\ET > 100 \GeV $ are required.
 This off-line selection is related to the jet plus missing $\ET$ trigger.  
The missing $\ET$  is calculated from the energies deposited on the calorimeters and the muon momenta. 
If the long-lived stau passes through the muon chamber 
and it is fast enough to be observed as a muon ($\beta>0.8$), 
it is treated as a visible particle.
 \item $|\eta(\stau)| < 0.63$.
 \item $\PT(\stau) > 100\GeV$.
 \item The stau should decay in the TRT, but before the 3rd module, i.e.~$|z|<712\un{mm}$ and $563\un{mm}
 <R<863\un{mm}$.
 \item The kink angle $0.1 < \Delta\phi < \pi/2$, where $\Delta\phi$ is the difference of the azimuthal
 angles between $\stau$ and the charged daughter particle.
 \item The charged daughter particle ($e$ or $\mu$) should be reconstructed as a track. To this purpose, at least it should fly through the outermost layer of the barrel TRT (i.e.~reaches $R = 1066\un{mm}$ satisfying $|z|<712\un{mm}$). The reconstruction efficiency for it is set to be 0.7 if $\PT>20\GeV$, or 0.6 if $>10\GeV$; otherwise it fails this selection.
\end{enumerate}

When the daughter particle is a tau lepton, the above selection strategy should be modified slightly because the daughter tau decays promptly. The tau decays into leptons or hadrons, and we take
both cases into account for the analysis. Especially, when the tau decays hadronically, it can generate
multiple charged particles. It is noteworthy that, since the tau is highly boosted in the $\stau$ decay,
they fly into a relatively narrow conical region. Thus, we replace the last two items above by the following two conditions.
\begin{enumerate}[1$'$.]
\setcounter{enumi}{4}
 \item The ``daughter cone'' (see below) should be separated away from the track of the mother particle by
 $0.1<\Delta\phi<\pi/2$ with respect to the azimuthal angle.
 \item Among decay products of the tau, at least one charged particle should be reconstructed. The criterion is the same as the previous cut~6.
\end{enumerate}
Here a ``daughter cone'' is defined as the smallest cone which covers all the tracks of the charged
daughter particles of the tau. Note that in the case of the 1-prong tau decay, the cone corresponds
to the charged daughter particle, and the separation condition 5$'$ is reduced to the kink
angle condition 5.

\begin{table}[t]
\begin{center}
\caption{%
The cut flow for $c\tau=400\un{mm}$ at the BC1 mass spectrum.
The number of events is displayed in the group I, III and III$'$, while the number of the
stau is shown in II and II$'$. The daughter particle is supposed to be $e/\mu$ in II and III.
When the daughter particle is $\tau$, the last two items of II are replaced by II$'$,
and the result becomes III$'$.
Note that one event contains two staus.
The reconstruction efficiency of daughter track is taken into account.
``Triggered'' means the number after the cut~1,
and the trigger efficiency ($>90\%$) is not included.
}
\label{tab:Summary}
\par
\begin{tabular}[t]{|c|l@{\,}c|r@{\ }l|r@{\ }l|r@{\ }l|}\hline
&&
&\multicolumn{2}{|c|}{$7\TeV,   2\invfb$}
&\multicolumn{2}{|c|}{$8\TeV,   5\invfb$}
&\multicolumn{2}{|c|}{$14\TeV, 10\invfb$}\\\hline\hline
I     & total SUSY event                &         &\ 673& events &\ 2832& events &\ 42463& events\\\cline{2-9}
      & triggered event                 &(cut~1)   & 426 & events & 1938 & events & 36240 & events\\\hline\hline
II    &  $\stau$ track                  &          & 852 &        & 3876 &        & 72480 &       \\\cline{2-9}
      & $|\eta(\stau)|<0.63$            &(cut~2)   & 409 &        & 1748 &        & 28535 &       \\\cline{2-9}
      & $\PT(\stau)>100\GeV$            &(cut~3)   & 378 &        & 1641 &        & 26642 &       \\\cline{2-9}
      & $\stau$ decay in TRT 1st/2nd    &(cut~4)   &  67 &        &  230 &        &  3642 &       \\\cline{2-9}
      & kink $0.1<\Delta\phi<\pi/2$     &(cut~5)   &  46 &        &  179 &        &  2601 &       \\\cline{2-9}
      & daughter reconstructed          &(cut~6)   &  28 &        &  101 &        &  1586 &       \\\hline\hline
III   & event with 1 or 2 kink          &          &  24 & events &  100 & events &  1563 & events\\\cline{2-9}
      & event with 2 kinks              &          &   4 & events &    1 & event  &    23 & events\\\hline
\multicolumn{7}{l}{}\\[-8pt]
\multicolumn{7}{l}{for the case where the stau emits $\tau$:}\\[7pt]\hline
II$'$ & separation $0.1<\Delta\phi<\pi/2$&(cut~5$'$)&  52 &        &  189 &        &  2805 &       \\\cline{2-9}
      & daughter reconstructed          &(cut~6$'$)&   26 &        &   95 &        &  1391 &       \\\hline\hline
III$'$& event with 1 or 2 kink          &          &   24 & events &   92 & events &  1374 & events\\\cline{2-9}
      & event with 2 kinks              &          &    2 & events &    3 & events &    17 & events\\\hline
\end{tabular}
\end{center}
\end{table}

The numbers of the remaining SUSY events (or stau tracks in group II)  are summarized in Tab.~\ref{tab:Summary},
in which decay length $c\tau$ of stau is assumed to be $400\un{mm}$.
This decay length corresponds to $\lambda\s{eff}=1.3\EE-8$ for the R-parity violation models,
and $m_{3/2}=0.35\keV$ for the gravitino LSP models, for the stau mass $m_{\stau}=140\GeV$ in the present model.
(See Eqs.~(\ref{eq:ctaugravitino}) and (\ref{eq:ctauRpV}).)
The selections on the staus are employed in the group  II (or II$'$ when a tau lepton is emitted by the stau);
the last numbers of the group II  denote the numbers of the stau kinks at various center of mass energies.
Finally in the group  III (or III$'$), we show the number of the kink events, that is,
events with at least one stau kink, and with two kinks.

We find that, in this value of $c\tau$, several percent of the total SUSY events could be detected.
In particular, the requirement that the stau should decay in the barrel TRT (the 4th item of the group  II)
significantly reduces the events.   
We can enlarge the acceptance by including the staus decaying in the end-cap TRT, but not yet performed in this note.
Even using only barrel TRT, the number of the kink events remains 
$\Order(1\%)$ of the total SUSY events.

As we discussed, when the daughter charged particle is a tau lepton, we slightly modify
the selection strategy.
However, to our delight, most of the daughter cones have small opening angles because of the boost;
it was checked that the peak of the azimuthal opening angle distribution sits around $0.01\un{rad}$ and
almost all are $<0.1\un{rad}$.
Therefore, the result appears very similar to the case where the daughter leptons are $e$ or $\mu$.

\begin{figure}[p]
 \begin{center}
  \includegraphics[width=370pt]{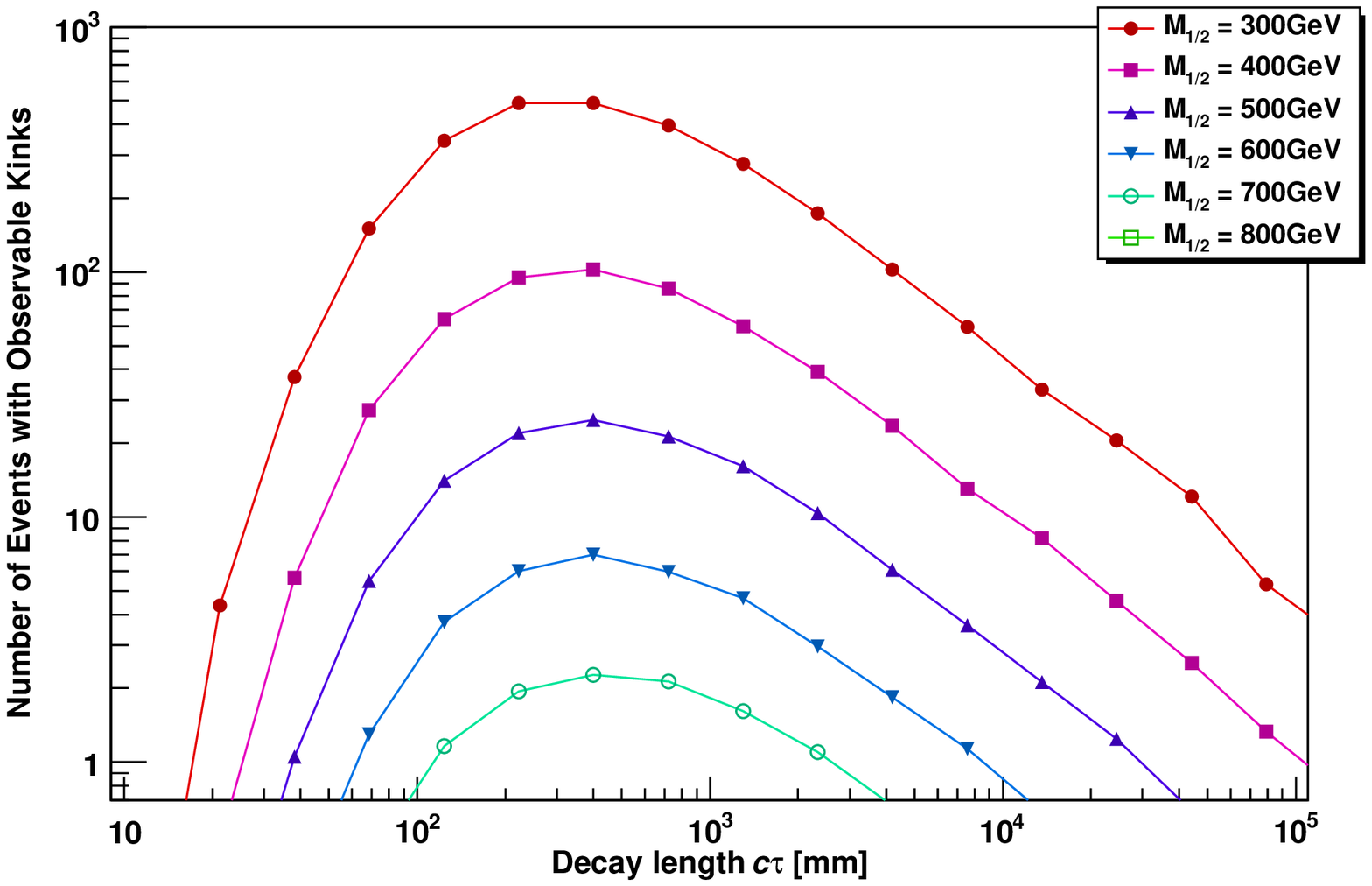}
  \caption{%
Dependence of the number of the kink events on the stau decay length $c\tau$ for $8\TeV$, $5\invfb$.
We also varied $M_{1/2}$, one of the parameters of the mSUGRA model; the other parameters are the
same as Tab.~\ref{tab:SUSYparams}. 
The daughter particle is assumed to be an electron, though the result does not change if it is a muon or a tau lepton.
In the simulation we generate 200000 events for each point to reduce the statistical uncertainty.
}
  \label{fig:length_dep_8TeV_num}
 \end{center}

 \begin{center}
  \includegraphics[width=370pt]{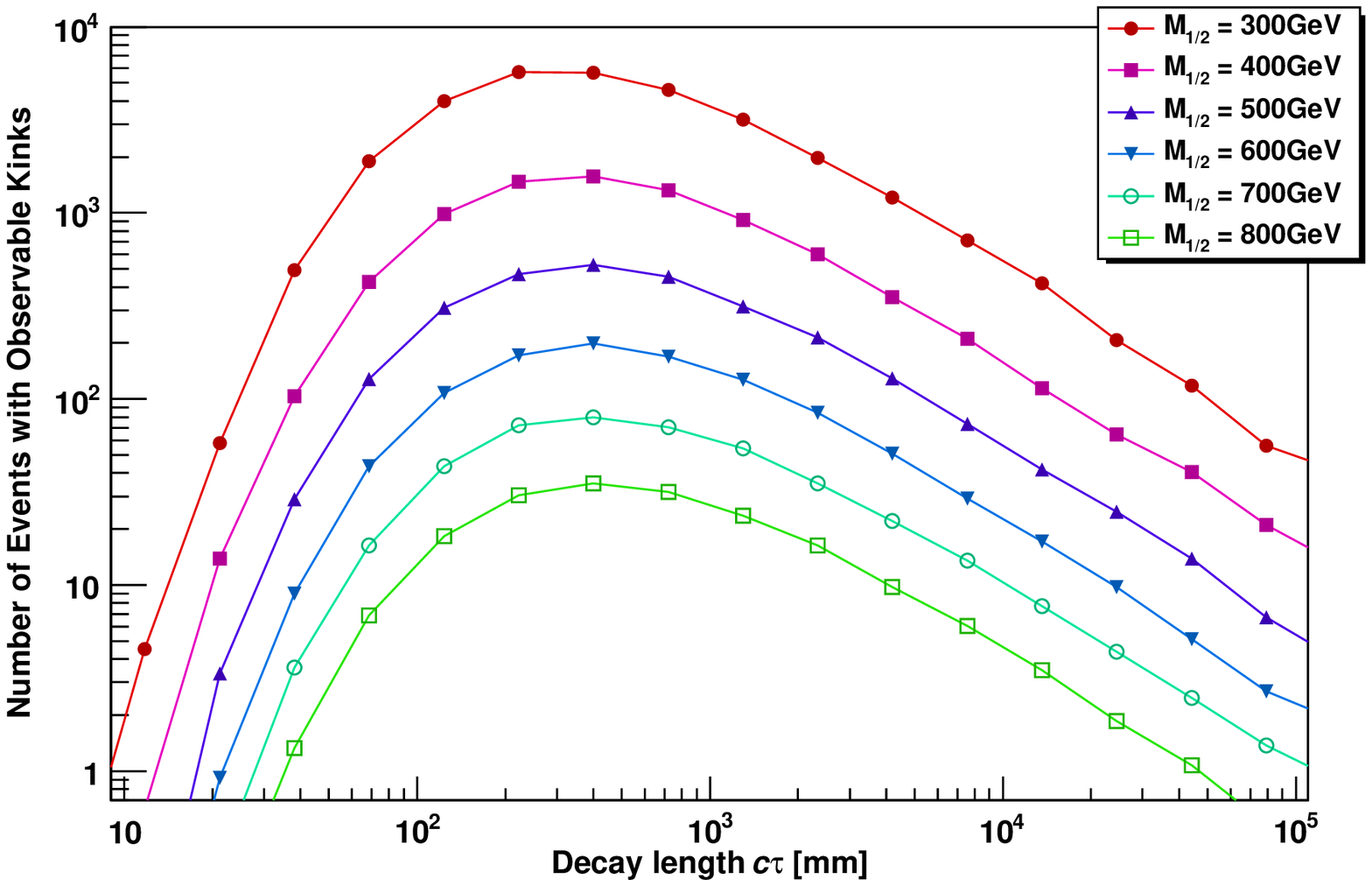}
  \caption{The same as Fig.~\ref{fig:length_dep_8TeV_num} but for $14\TeV$, $10\invfb$.}
  \label{fig:length_dep_14TeV_num}
 \end{center}
\end{figure}

\begin{table}[t]
\begin{center}
\caption{%
The superparticle masses at the weak scale and the total SUSY cross section for the model points considered
in Fig.~\ref{fig:length_dep_8TeV_num}.
We vary only $M_{1/2}$, and other parameters are the same as Tab.~\ref{tab:SUSYparams}.
}
\label{tab:MASSandCS}	
\catcode`?=\active \def?{\phantom{0}} \catcode`@=\active \def@{\phantom{.}}
\begin{tabular}[t]{|c|c|c|c|c|}\hline
$M_{1/2}$
& \multicolumn{2}{|c|}{Masses (GeV)}
& \multicolumn{2}{|c|}{Cross section (fb)}\\\cline{2-5}
(GeV) & \kern11pt$\stau$\kern11pt & \kern11pt$\tilde g$\kern11pt & $8\TeV$ & $14\TeV$\\\hline
300  & 103  & ?715  & $2.95\EE3$ & $1.81\EE4$  \\\hline
400  & 140  & ?932  &  556@??    & $4.22\EE3$  \\\hline
500  & 176  & 1145  &  143@??    & $1.31\EE3$  \\\hline
600  & 212  & 1355  &  ?44.5?    &  472@??     \\\hline
700  & 248  & 1562  &  ?17.8?    &  194@??     \\\hline
800  & 283  & 1768  &  ??6.12    &  ?87.1?     \\\hline
\end{tabular}
\end{center}
\end{table}

In Fig.~\ref{fig:length_dep_8TeV_num}, we present the dependence of the number of the kink
events on the stau decay length $c\tau$ for $8\TeV$, $5\invfb$ setup.
Here, the daughter particle is assumed to be an electron,
but the result does not change if it is a muon or a tau lepton.
The figure tells that the number greatly depends on the decay length.
This is because we demand that staus should decay in a very restricted region.

In the figure, we also changed the SUSY parameter $M_{1/2}$.
We show the masses of the relevant superparticles and the SUSY total cross section in Tab.~\ref{tab:MASSandCS}.
{}From Fig.~\ref{fig:length_dep_8TeV_num} and Tab.~\ref{tab:MASSandCS},
we can see that the ratio of the final number of the events to the number of the total events, for a given $c\tau$, is independent of $M_{1/2}$.
In fact, the number of the kink events is mainly controlled by
the total SUSY cross section and the decay length $c\tau$.
Here we can see that, when the stau mass is $100 \TO 200\GeV$, the kink events can be discovered
in a wide region of the decay length.

We also show the same result for the $14\TeV$, $10\invfb$ LHC setup in
Fig.~\ref{fig:length_dep_14TeV_num}.
The $c\tau$ dependence is quite similar to the $7\TeV$ case, but the number significantly increases due to the large cross section.
If the decay length is $\Order(10\TO 10^5)\un{mm}$,
we can achieve the discovery of the kinks in a wide region of the SUSY parameters.

In Tab.~\ref{tab:range}, the accessible range of the lifetime are shown for
the setup $8\TeV$ $5\invfb$ and $14\TeV$ $10\invfb$.
We have also shown the corresponding ranges of the gravitino mass $m_{3/2}$ for the gravitino LSP scenario,
and that of the effective R-parity violating coupling $\lambda_{\rm eff}$ for the models with R-parity violation.
For $14\TeV$ $10\invfb$ ($8\TeV$ $5\invfb$), the ranges are
$0.03\keV\lsim m_{3/2}\lsim 8\keV$ ($0.04\keV\lsim m_{3/2}\lsim 2\keV$)
and
$4\EE{-10}\lsim \lambda_{\rm eff}\lsim 8\EE{-8}$ 
($1\EE{-9}\lsim \lambda_{\rm eff}\lsim 6\EE{-8}$),
respectively.


\section{Discussion/Conclusion}
\label{sec:discuss}

In this paper, we have investigated the possibility of probing SUSY models
by analyzing the kink signatures of charged tracks
in the TRT of the ATLAS detector.
As underlying models, we discussed two models; (a) SUSY models with a gravitino LSP and a stau NLSP, and (b) R-parity violating SUSY models with a stau (N)LSP.
It was shown that, if the decay length is $\Order (10 \TO 10^5)\un{mm}$,
a large number of kink events can be discovered in a wide range of the SUSY parameters.
For relatively light SUSY particles, i.e., gluino mass $\lsim 1\TeV$  and stau mass $\lsim 200\GeV$, 
kinks can be observed already in the early stage of the LHC.

In the case of gravitino LSP scenario, the kink signature can be observed
for $\Order(0.01\keV \TO 1\keV)$ (see Tab.~\ref{tab:range}).
As discussed in Sec.~\ref{sec:models-with-grav}, 
this corresponds to the most challenging gravitino mass range from 
cosmological point of view~\cite{gravitino_keV,Feng:2010ij}.
The reheating temperature after inflation must be $\Order(100\GeV)$ or lower
unless there is a large entropy production.
To put it the other way around, if such a kink signature is indeed observed
and if it becomes likely that the underlying theory is a gauge-mediated SUSY breaking model with the gravitino LSP,
it has a significant impact on the cosmology.
In particular, both of the inflation and the baryogenesis should occur at a very low energy scale.

\begin{table}[t]
\begin{center}
\caption{%
The accessible range of decay length (more than 10 kinks can be observed)
and the corresponding ranges of R-parity violating coupling
and the gravitino mass for the models in Tab.~\ref{tab:MASSandCS},
at $8\TeV\ 5\un{fb^{-1}}$ and $14\TeV\ 10 \un{fb^{-1}}$,
}
\label{tab:range}

\catcode`?=\active \def?{\phantom{0}} \catcode`@=\active \def@{\phantom{.}}
\begin{tabular}[t]{|c|c|c|c|c|c|}
\hline
$M_{1/2}$
& \multicolumn{2}{|c|}{Masses (GeV)}
& \multicolumn{3}{|c|}{$N>10$ at $8\TeV$ $5\invfb$}
\\\cline{2-6}
(GeV)
& \kern11pt$\stau$\kern11pt
& \kern10pt$\tilde g$\kern10pt
& decay length (mm)
& $m_{3/2}$ (keV)
& $\lambda_{\rm eff}$
\\
\hline
300  & 103  & ?715 & ?30 \ -- \ $5\EE4$ & 0.04 \ -- \ 2 & $(\,0.1 \ \TO\  6\,)\EE{-8}$  \\ \hline
400  & 140  & ?932 & ?50 \ -- \ $1\EE4$ & 0.1? \ -- \ 2 & $(\,0.3 \ \TO\  4\,)\EE{-8}$  \\ \hline
500  & 176  & 1145 & 100 \ -- \ $2\EE3$ & 0.3? \ -- \ 2 & $(\,0.5 \ \TO\  2\,)\EE{-8}$  \\ \hline
\multicolumn{6}{c}{}\\[-5pt] \hline
$M_{1/2}$
& \multicolumn{2}{|c|}{Masses (GeV)}
& \multicolumn{3}{|c|}{$N>10$ at $14\TeV$ $10\invfb$}
\\\cline{2-6}
(GeV)
& \kern11pt$\stau$\kern11pt
& \kern10pt$\tilde g$\kern10pt
& decay length (mm)
& $m_{3/2}$ (keV)
& $\lambda_{\rm eff}$
\\
\hline
300  & 103  & ?715  &  ?10 \ -- \ $5\EE5$ &  0.03 \ -- \ 6 & $(\,0.04\, \TO\  8\,)\EE{-8}$  \\ \hline
400  & 140  & ?932  &  ?20 \ -- \ $2\EE5$ &  0.08 \ -- \ 7 & $(\,0.07\, \TO\  6\,)\EE{-8}$  \\ \hline
500  & 176  & 1145  &  ?30 \ -- \ $6\EE4$ &  0.2? \ -- \ 7 & $(\,0.1?\, \TO\  4\,)\EE{-8}$  \\ \hline
600  & 212  & 1355  &  ?40 \ -- \ $2\EE4$ &  0.3? \ -- \ 8 & $(\,0.1?\, \TO\  3\,)\EE{-8}$  \\ \hline
700  & 248  & 1562  &  ?60 \ -- \ $1\EE4$ &  0.5? \ -- \ 7 & $(\,0.2?\, \TO\  3\,)\EE{-8}$  \\ \hline
800  & 283  & 1768  &  ?90 \ -- \ $4\EE3$ &  0.9? \ -- \ 7 & $(\,0.3?\, \TO\  2\,)\EE{-8}$  \\ \hline
\end{tabular}
\end{center}
\end{table}


When the R-parity violation is a source of the kink track generation, the trilinear coupling is in the 
range of $\Order(10^{-10} \TO 10^{-8})$. Although the MSSM-LSP can not be a dark 
matter, the gravitino LSP is a viable dark matter candidate. In this case, 
the severe constraint 
from the big-bang nucleosynthesis due to the superparticle decays can be avoided because of the R-parity violation.
If the gravitino has a mass of $\sim 10\GeV$, the thermal leptogenesis can work 
successfully to generate the baryon asymmetry of the universe without suffering from the baryon 
washout.
The decay of the gravitino might be detected by measurements of the cosmic rays, 
depending on the gravitino mass as well as the pattern of the R-parity violating parameters.

Once the kink signature is discovered, the next step will be 
the discrimination of the kink models.
Characteristics of the model are embedded in the final state particles of the kink decay.
For instance, if we turn only on the coupling $\lambda_{i31}$ in the R-parity violating models, the
final state is an electron (see Tab.~\ref{tab:DaughterFractions}). For $c\tau = 400\un{mm}$,
it is expected from Tab.~\ref{tab:Summary} that we observe $\sim 20$ electron events
for the LHC at $7\TeV$, $2\invfb$, and  $\sim 100$ for $8\TeV$, $5\invfb$ at the BC1
benchmark mass spectrum. At the $14\TeV$ run, the number increases up to $\sim 1000$
for the integrated luminosity $10\invfb$. On the other hand, when we switch on $\lambda_{i32}$
instead of $\lambda_{i31}$, the final state becomes a muon, which can be distinguished from
the electron events by the detector signatures. Thus, it is expected that these models can be discriminated
even in the early stage of the LHC run and much better for $14\TeV$. In the case of the tau
final state, e.g.~in the gravitino LSP case, $\sim 20$ $\tau$-jet events could be observed
in the early LHC, while the numbers of the $e$ and $\mu$ events are at most a few.
In the $14\TeV$ run, the number significantly increases, which enables us to distinguish
the $e, \mu$ and $\tau$ final-state models.

In the case of the R-parity violating scenarios, our study covers the $L_iL_j\bE_k$ interactions and
generally, the bilinear $L_i\Hu$ terms.
The $L_iQ_j\bD_k$ interactions, as well as some case with the bilinear violation,
make staus decay into two jets plus 0 or 2 lepton(s),
which are not analyzed in this paper.
The stau decaying hadronically in detectors is an interesting subject to study,
which would complement our work.

\section*{Acknowledgment}
The authors are grateful to Takeo Moroi for useful comments and careful reading of the manuscript,
and thank Satoshi Shirai and Kouhei Nakaji for useful comments.
The work of K.H. was supported by JSPS Grant-in-Aid for Young Scientists (B) (21740164)
and Grant-in-Aid for Scientific Research (A) (22244021).
The work of S.I. was supported by JSPS Grant-in-Aid for JSPS Fellows.
This work was supported by World Premier International Center Initiative (WPI Program),
MEXT, Japan.
{\small

\begingroup\raggedright\endgroup
}
\end{document}